\documentclass[journal]{IEEEtran}
\usepackage{cite}

\usepackage[font=small,labelfont=bf]{caption}

\ifCLASSINFOpdf
\else
\fi
\usepackage{caption}
\usepackage{graphicx}
\usepackage{subcaption}
\DeclareCaptionType{equ}[][]
\usepackage[ruled, vlined, noalgohanging]{algorithm2e}
\usepackage{mathtools}
\usepackage{amsthm}
\usepackage{xcolor}
\newtheorem{remark}{Remark}
\newtheorem{definition}{Definition}

\usepackage{url}
\usepackage{booktabs}
\usepackage{calrsfs}
\usepackage{orcidlink}

\usepackage{xspace}
\newcommand{\ModelOne}{\textsc{OTSP-Model-1}\xspace}
\newcommand{\ModelTwo}{\textsc{sDCOPF-Model-2}\xspace}
\newcommand{\ModelThree}{\textsc{rOTSP-Model-3}\xspace}
\newcommand{\ModelFour}{\textsc{hOTSP-Model-4}\xspace}
\newcommand{\ModelFive}{\textsc{ACOPF-Model-5}\xspace}

\newcommand{\lpsc}{\textsc{lpsc}\xspace}
\newcommand{\AlgOne}{\textsc{Algorithm 1}\xspace}
\newcommand{\AlgTwo}{\textsc{Algorithm 2}\xspace}
\newcommand{\AlgThree}{\textsc{Algorithm 3}\xspace}

\newcommand{\dcopf}{\textsc{sDCOPF}\xspace}
\newcommand{\rotsp}{\textsc{rOTSP}\xspace}

\newcommand{\opf}{\textsc{OPF}\xspace}
\newcommand{\otsp}{\textsc{OTSP}\xspace}
\usepackage[bb=boondox]{mathalfa}

\newcommand{\V}{\mathcal{V}}
\newcommand{\G}{\mathcal{G}}
\newcommand{\E}{\mathcal{E}}
\newcommand{\C}{\mathcal{C}}
\newcommand{\Es}{\mathcal{E}^{S}}
\newcommand{\Ea}{\mathcal{E}^{A}\xspace}
\newcommand{\Ei}{\mathcal{E}^{I}\xspace}
\newcommand{\J}{\mathcal{J}\xspace}

\newcommand{\lp}{\mathbb{L}\xspace}
\newcommand{\gap}{\textit{gap}\xspace}
\newcommand{\ub}{\textit{ub}\xspace}
\newcommand{\lb}{\textit{lb}\xspace}
\newcommand{\case}{\mathbb{c}\xspace}
\newcommand{\xn}{\boldsymbol{x}\boldsymbol{0}\xspace}
\newcommand{\inc}{\textit{inc.}}
\newcommand{\stat}{\textit{stat.}}
\newcommand{\sol}{\textit{sol.}}
\newcommand{\bld}[1]{\boldsymbol{#1}}

\newcommand{\indf}{\boldsymbol{1}_{\Ea}\xspace}

\usepackage{float}
\floatstyle{ruled}
\newfloat{model}{thp}{lop}
\floatname{model}{\textsc{Model}}

\SetCommentSty{mycommfont}

\begin{document}
%

\title{Optimal Transmission Switching: \\Improving Exact Algorithms by \\ Parallel Incumbent Solution Generation}
\title{Optimal Transmission Switching: \\Improving Exact Algorithms Using Heuristics}
\title{Optimal Transmission Switching: \\Improving Exact MILP Algorithms Using Heuristics}
\title{Optimal Transmission Switching: \\Improving Solver Performance Using Heuristics}

%
%
%

\author{Anton~Hinneck~\orcidlink{0000-0003-1929-5366},~\IEEEmembership{Student Member,~IEEE},~%
        David~Pozo~\orcidlink{0000-0002-4080-1828},~\IEEEmembership{Senior Member,~IEEE}
}

\maketitle

\begin{abstract}
The optimal transmission switching problem (OTSP) is an NP-hard problem of changing the topology of a power grid to obtain an improved dispatch by controlling the operational status of the transmission lines. Exact solution techniques based on mixed-integer programming (MIP) like branch-and-bound (B\&B) guarantee identifying global optimal solutions if they exist but are potentially intractable in realistic power grids. Heuristic methods, on the other hand, can provide tractable solution approaches but potentially cut off optimal solutions. Heuristics are implemented along with B\&B in modern MIP solvers to take advantage of both approaches. 

This paper proposes solving the full OTSP formulation alongside parallel heuristics that generate good candidate solutions to speed up conventional B\&B algorithms. The innovative aspect of this work is a new asynchronous parallel algorithmic architecture and the exploitation of domain-specific knowledge in parallel heuristics. Heuristics generate solutions asynchronously to be injected into the full OTSP solution procedure during run time.  

Our method is tested on 14 instances of the pglib-opf library: The largest problem consists of 13659 buses and 20467 branches. Our results show good performance for large problem instances, with consistent improvements over off-the-shelf solver performance. 

\end{abstract}

\begin{IEEEkeywords}
Optimal transmission switching problem (OTSP), high-performance computing (HPC), topology control, mixed-integer linear programming (MILP)
\end{IEEEkeywords}

%
\IEEEpeerreviewmaketitle

\section{Introduction}

\subsection{Motivation}
\IEEEPARstart{T}{ransmission} infrastructure needs to be operated continuously whilst meeting high security standards to protect against unforeseen events and disturbances. Managing grids of real-world scales requires high-performance algorithms for computationally expensive problems to be solved. N-1 security is widely adopted in the industry, e.g. by the American Electric Reliability Corporations
(NERC) that require to ``\textit{demonstrate through a valid assessment that \ldots the interconnected transmission system \ldots can be operated to supply projected customer demands and projected \ldots transmission services, at all demand levels over the range of forecast system demands, under the contingency conditions''} \cite{nerc2010} that a single generator, transmission line or transformer might fail. To hedge against failures, system operators employ preventive and corrective measures. Preventive measures are taken before a malfunction occurs, while corrective measures are taken after the fact. Several corrective measures are used in the industry. The PJM for instance, depending on the state of the system, approves predetermined switching actions as non-cost responses alongside redispatch, curtailment, or load shedding in emergency situations \cite{pjm2021}. Likewise, switching transmission elements is an approved control action by the Independent System Operators of New England (ISO-NE) \cite{isone2021}.

Transmission switching, which considers switching transmission lines, was identified as an effective control mechanism for electricity grids, to reduce losses, improve security, and control short-circuit currents \cite{Glavitsch1985}.
It also poses an effective measure to relieve system overloads \cite{Mazi1986}.
Although switching actions have been used for decades in the industry to prevent voltage violations and improve voltage profiles \cite{Shao2005}, specifically improving economic objectives is a more recent occurrence. Using switching, as a control action, comes at no additional costs compared to load shedding or curtailment when facing overloads.

\subsection{Literature review and open challenges}

Employing a system's switching capabilities to improve the economic dispatch was introduced by Fisher et al. \cite{Fisher2008}.
However, switching lines raises concerns about reliability of the system \cite{Hedman2009}. 
Hence, the addition of security constraints is identified to be of high importance \cite{Hedman2009}, \cite{Liu2012}.
Other extensions were developed as well, taking into account multiple periods \cite{Liu2012}, expansion planning \cite{Khodaei2010-1}, investments \cite{Villumsen2012}, and security-constrained unit commitment \cite{Khodaei2010-1}. Recent work considers contingencies, unit commitment, generation scheduling, and uncertainty simultaneously \cite{Saavedra2020}. 
However, one of the main issues is the computational expense of the problem, as stated in \cite{Fisher2008,Saavedra2020}.
This especially applies for systems with hundreds or thousands of lines.
The optimal transmission switching problem (\otsp) was introduced in \cite{Fisher2008} as a \textit{mixed-integer linear formulation} (MILP). Such problems can, in theory, be solved with methods that guarantee to find an optimal solution, if it exists.
Obtaining a solution to the \otsp can easily become intractable because the problem was proven NP-hard \cite{Lehmann2014} and real-world grids can be of significant size.
Related versions of the \otsp and methods proposed in literature are usually tested on one or two test cases \cite{Hedman2009}, \cite{Fuller2012}, \cite{Fattahi2019}, \cite{Barrows2012}, which rarely exceed the IEEE 118-bus case's size regarding the amount of binary variables.
\textit{Fattahi et al} \cite{Fattahi2019} solved several problem instances from the \verb+pglib-opf+ library with up to 400 binary variables per model, i.e. 400 switchable lines. The computational burden becomes much more significant when additional extensions such as contingencies, unit commitment and multiple periods are considered in addition to switching. 
The work in \cite{Hedman2010} and \cite{Ayala2014} are two of the few contributions we found that consider all possible switching actions. This results in high computation times, high optimality gaps, and only small test cases being tractable. Most other contributions apply heuristic methods, limiting the number of simultaneously deactivated lines using a cut or a reduced set of switchable lines from the start, to yield tractable problems \cite{Fisher2008}, \cite{Hedman2008}, \cite{Hedman2009}, \cite{Saavedra2020}.
Switching problems as stated in \cite{Fisher2008} are integral part in many modern approaches. Being able to obtain better solutions quicker could be a valuable contribution in several fields:
\paragraph{Robust optimization and scenarios}\label{sec:rob} When operating a power system maintaining its stability is crucial. Let a set of contingencies be denoted by $\C$. Many security-constrained problems decompose into a structure similar to \eqref{eq:rob}. Here, $\boldsymbol{y}$ represent the decision variables in the first stage, $\C$ a set of scenarios or contingencies, and $\boldsymbol{x}$ the recourse variables. Recourse variables can be understood as corrective measures, after a contingency occurred, i.e. uncertainty is revealed.
\begin{equation}
\label{eq:rob}
    \underset{\boldsymbol{y}}{\mathrm{min}}~\underset{c \in \C}{\mathrm{max}}~\underset{\boldsymbol{x}}{\mathrm{min}}
\end{equation}
The N-1 security criterion adopted by PJM is wide-spread and results in large-scale MIPs where potentially many second-stage switching decisions must be made, even using advanced methods like column-and-constraint generation (CCGA) \cite{Saavedra2020}, \cite{Ayala2014}, \cite{Zeng2013}, \cite{Monticelli1987}. 
\paragraph{Bus splitting and optimal substation reconfiguration}
OTS only considers the switching of transmission lines. However, more complex control schemes were proposed. Bus splitting \cite{Shao2005} models the separation of bus bars within substations. Optimal substation reconfiguration (OSR) as proposed in \cite{Nasrolahpour2012}, \cite{Heidarifar2016}, \cite{Hinneck2021} offers the highest degree of flexibility, trying to consider all possible switching actions in substations.
Whilst higher decreases in congestion have been reported than could be achieved by OTS alone \cite{Hinneck2021}, these models increase the computational burden significantly.
However, it has been shown that any solution to OTS is a solution for OSR. This applies to the node-breaker models proposed in \cite{Heidarifar2016} and \cite{Hinneck2021}, specifically. Devising a method that yields high-quality OTS solutions thus yields high-quality OSR solutions as well.%
\paragraph{Alternating current and power markets} Optimal power flow models are one of the most important problems system operators must solve on a rolling basis. While the exact physical power flow equations are highly complex for alternating current, approximations like the linear DC approximation allow to derive market signals based on duality theory. Moreover, highly developed solvers are available for MILPs. Linear DC solutions are commonly used to find AC-feasible operating points. AC-feasiblity is vital to monitor the physical system state \cite{ONeill2012},\cite{Castillo2013I},\cite{Castillo2013II}. Switching actions accounting for AC power flow equations have therefore long been considered \cite{Henneaux2016}, \cite{Khanabadi2013}, \cite{Barrows2014}, \cite{Poyrazoglu2015}, \cite{Capitanescu2014}, 
\cite{Xingpeng2017}, \cite{Crozier2022}.  We will close our experiments with an outlook on AC-feasibility of obtained topologies.

\color{black}
High-performance solution methodologies for problems remain to be critical in the current power systems literature.
Several \textit{heuristic methods} were proposed to find improved feasible solutions for the problem quickly, rather than finding an optimal one. These methods usually identify (a subset of) lines based on sensitivities \cite{Ruiz2011}, \cite{Fuller2012}, i.e. a reduction of the solution space by selecting promising candidate lines \cite{Barrows2012}. 
This is justified by the fact that a small selection of lines is responsible for most of the economic benefit \cite{Barrows2012}.
As good feasible but non-optimal solutions can be computed efficiently \cite{Fuller2012}, \cite{Barrows2012}, \cite{Pourahmadi2016}, \cite{Soroush2014}, \cite{Papavasiliou2013}, merely using heuristics might seem appropriate to improve the \opf, especially for real-world grids.
However, not using an exact algorithm to solve MILP's has an important disadvantage:  No lower bound on the full problem is available. The lower bound denotes the minimal objective value (cost of generation) that could be achieved if a corresponding integral solution existed. In other words, no feasible solution with a better objective value exists.
Without this information, the quality of a given solution cannot be assessed other than by its improvement over a known incumbent solution. Using lower bounds not only facilitates exact methods, but also poses a strong argument for the application of exact methods in real markets. How much improvement could still be possible, even if a problem can not be solved optimally in a given time, can be quantified.  
Moreover, modern MIP solvers implement general purpose heuristics which are competitive with heuristics proposed in the power systems literature, which can additionally be taken advantage of.
The above-mentioned drawbacks, namely, i) of solving the full MILP \mbox{formulations} on computation time and ii) of heuristic methods for not providing optimal solution guarantees motivate our work. We intend to combine exact algorithms implemented in MILP solvers (global solution guarantee) with heuristic methods (computational speed).
The work in this paper is enabled by rich APIs of commercial solvers and parallel computing frameworks.

\subsection{Branch-\&-bound algorithms and heuristics}

\subsubsection{A general exposition of modern B\&B algorithms}
Branch-and-bound (B\&B) algorithms pose the basis for all relevant MIP solvers and are exact solution techniques.
Rather than enumerating all possible solutions, B\&B algorithms implement systematic search, enumerating candidate solutions and eliminating impossible ones.
Branch-and-cut (B\&C) algorithms extend B\&B by incorporating the computation and enforcement of cutting planes to reduce the search space \cite{Young2015} into the process. 
Such methods can be employed to solve combinatorial optimization problems like transmission switching, if a MILP formulation exists.

For minimization problems, a \textit{lower bound} ($lb$) on a MILP can be obtained by solving a continuous relaxation of the original problem. If this relaxation has a solution, its objective value denotes the best objective value possible by any integer-feasible solution.
Obtaining an \textit{upper bound} ($ub$) (i.e. an integer-feasible solution for a MILP) can be difficult due to the complexity of many real-world problems \cite{Huang2009}.
The best feasible solution available at any moment of the iterative B\&B algorithm or one of its derivatives is called \textit{incumbent}. Once $ub$ and $lb$ are available, an \textit{optimality gap} can be computed as $\gap = 100 \frac{\left|ub - lb\right|}{lb}$. If $\gap\leq\epsilon$, where $\epsilon$ denotes a predefined tolerance, a solution is classified optimal and optimization terminates. 

In B\&B algorithms, a search tree is used to represent the solution space.
The underlying decision schemes vastly affect the convergence rate, such as selecting nodes to branch on and finding feasible upper bounds \cite{Huang2009}.
Even though B\&B algorithms are guaranteed to provide an optimal solution, in practice, this might not be feasible due to time constraints. Hence, heuristics that generate high-quality solutions quickly are one of the most important recent improvements in MIP solvers \cite{Fischetti2011}.
State-of-the-art solvers, like Gurobi, take advantage of heuristics to find feasible solutions in the original problem (and therefore \textit{ub}s). 
Quickly generated upper bounds facilitate extensive node pruning in the search tree and therefore reductions of the search space at large \cite{Fischetti2011}. As a measure of improvement, decreasing the primal integral is widely employed, i.e. reducing the area under the $ub$ for minimization problems \cite{Berthold2013}.
One type of heuristic that has proven powerful are MIP heuristics. These construct restricted versions of an original problem, which in turn can be solved with a MIP solver \cite{Fischetti2011}.
A powerful MIP improvement heuristics is relaxation-induced neighborhood search (RINS) \cite{Danna2005}. Improvement heuristics require knowledge of a feasible incumbent.
Methods like local branching do not require a known solution but construct a sub-MILP by adding cuts that cut off parts of the search space.
\subsubsection{Heuristics on transmission switching}
\label{sec:hts}
{Heuristic concepts mentioned in the previous subsection were developed for general MIPs but can also be found in the power systems literature, albeit in adjusted form. In \cite{Fisher2008}, \cite{Hedman2008} and \cite{Hedman2009} cuts are added to the full problem to make it tractable but potentially cut off good solutions.
In \cite{Fuller2012}, \cite{Liu2012b} and \cite{Papavasiliou2013} heuristics are utilized that are similar to neighborhood search as proposed in \cite{Danna2005}.}
For many problems in power systems, improvement heuristics are effective, as system operators generally know a feasible system state. This is implicitly taken advantage of in domain-specific work on switching heuristics like \cite{Fuller2012} and \cite{Ruiz2011}. The proposed methods start optimization with \textit{all lines active} to then remove candidate lines. {%
While previous work focuses on improvements, we want to devise a heuristic in this contribution that is applied while solving full problems, like implemented in off-the-shelf-solvers.
Whilst previous switching heuristic provide a good starting point, generalization of the work in \cite{Ruiz2011}, \cite{Fuller2012}, \cite{Liu2012b} and \cite{Papavasiliou2013} is required, to develop a method that can be applied in parallel and in conjunction with an exact algorithm like B\&C.}

\subsection{Problem statement and paper contributions}
\label{sec. requ}
In this paper, a parallel algorithmic architecture to solve MILPs like \otsp is proposed. We investigate how the asynchronous provision of good-quality solutions affects MIP solver performance in a master-subproblem framework. The main processor solves the full formulation of the \otsp (also referred to as the master problem), while a single or multiple worker instances solve a series of restricted sub-MILPs (subproblems). Subproblems consider a reduced solution space of topological control actions and can therefore be solved faster. Optimal solutions might be cut off, however.

Based on prior work, we formulate four main \textit{requirements}, based on which a heuristic is developed in this paper:

\paragraph{Arbitrary initial topologies}
Previous methods start optimization from a topology where all lines are active. We want our heuristic to be able to start a search from the best currently known topology. 
\paragraph{Convergence guarantees}
{We want to devise a method that yields restricted heuristic problems of arbitrary (but easily controllable) complexity and accuracy. These two terms refer to the retention level of the full problem's search space using a heuristic. The more search space is retained, the more accurately it reproduces the full problem and might contain good or optimal solutions. Simultaneously, a larger search space makes a respective (heuristic) problem more complex as more candidate solutions might have to be enumerated. We  provide a convergence guarantee in the proposed iterative master-subproblem frameworks along with the iterative process.}
\paragraph{Reversible switching decisions}
Most heuristic methods make permanent changes to the grid's topology. We require, however, that if adding a previously removed candidate line yields an improved system state, switching decisions can be reverted.  This makes MIP heuristics most viable, as opposed to greedy methods for example.
\paragraph{Scalability}
Modern hardware allows for parallel computations using multi-core or even multiple CPUs, which has been successfully applied in previous work on switching heuristics \cite{Papavasiliou2013}. We want to develop a heuristic considering that potential explicitly.
\newline
Contrary to the state-of-the-art algorithms using problem decomposition techniques applied in integer programming like Bender's decomposition, column-and-constraint-generation, and other cutting plane methods, which require solving the master and subproblems sequentially \cite{Trespalacios2014}, \cite{ Saavedra2020}, optimization of our master problem does not halt. It receives solutions from subproblems as they are produced. Every such obtained solution is feasible in the monolithic formulation. Moreover, these methods might produce an infeasible solution in the full problem without assumptions like full recourse.

The \textit{contributions} of this paper are three-fold:
\begin{itemize}
    \item \textit{Model formulation and properties:} We propose a method of restricting the \otsp to form sub-MILPs of arbitrary complexity and accuracy. We use domain-specific switching indicators to generate sub-MILPs, which in turn are used to produce solutions of the master problem.
    \item \textit{Parallel MIP heuristic:} We describe a parallel algorithm that, in the main process, solves the full \otsp. On a parallel process, it continuously generates and solves sub-MILPs. Rather than solving sub-MILPs sequentially, we send incumbents to the main parallel solver instance and receive its solutions asynchronously.
    \item \textit{Analytics and benchmarks:} {We have thoroughly inspected 14 test cases with up to 13659 buses and 20467 lines (binary variables), compared several heuristics and switching indicators.}
\end{itemize}

The rest of the paper is organized as follows: Section \ref{sec:math_framework} introduces models and their properties to create sets of switchable lines, construct and solve restricted problems and exchange feasible solutions, while section \ref{sec:algorithm_arqu} describes the algorithmic structure and parallel implementation. Large numerical analyses over 14 case studies are described in section \ref{sec:numerical_results}. Finally, results are discussed and conclusions are drawn in section \ref{sec:conclusion}.

\begin{table}[ht!]
  \caption{Nomenclature}
  \label{tab:notation}
\begin{center}
     \begin{tabular}{ r l l }
        \textbf{Symbol} & \textbf{Description}\\
        \toprule
        \multicolumn{2}{r}{\textit{Indexes and Sets}}\\
        \midrule
        $v, w \in \V$ & Buses\\
        $e \in \E$ & Transmission lines\\
        $g \in \G$ & Generators\\
        $c \in \C$ & Contingencies\\
        $j\in\J$ & Parallel worker ranks\\
        \midrule
        \multicolumn{2}{r}{\textit{Subsets \& Set Elements}}\\
        \midrule
        $v^{f}_{e}$/$v^{t}_{e} $ & From/To bus of line $e$\\
        $\E^{f}_{v}$/$\E^{t}_{v}$ & Transmission lines from/to at bus $v$ \qquad \qquad\\
        $\Ea/\Ei$ & Always-active/Always-inactive transmission lines\\
        $\Es$ & Selected transmission lines from $\lp$\\
        $\lp$ & Sorted list of indexes of all lines \\
        \midrule
        \multicolumn{2}{r}{\textit{Variables}}\\
        \midrule
        $\theta_{v}$ & Voltage angle at bus $v$ \hfill [rad]\\
        $p_{g}$ & Power generation of generator $g$ \hfill [p.u.]\\
        $f_{e}$ & Power flow on transmission line $e$ \hfill [p.u.]\\
        $x_{e}$ & Switching status of line $e$\\
        \midrule
        \multicolumn{2}{r}{\textit{Parameters}}\\
        \midrule
        $d_{v}$ & Demand at buses \hfill [p.u.]\\
        $b_{e}$ & Susceptance of line $e$ \hfill [p.u.]\\
        $\overline{f}_{e}$ & Maximal transmission capacity \hfill [p.u.]\\
        $\underline{p}_{g},~\overline{p}_{g}$ & Minimal/Maximal generation \hfill [p.u.]\\
        $\underline{\theta}_{v},~\overline{\theta}_{v}$ & Minimal/Maximal phase voltage angle \hfill [rad] \\
        $c_{g}(p_g)$ & Generation costs function \hfill [\$/p.u.]\\
        $M_{e}$ & Sufficiently large parameters (big M)\\
        $n$ & Number of lines in $\Es$\\
        $\Delta{n}$ & Number of new lines in $\Es$ per iteration\\
        \midrule
        \multicolumn{2}{r}{\textit{Dual variables}}\\
        \midrule
        $\alpha_{e}$ & Line profit switching criterion\\
        $\pi$ & Lmps\\
        \bottomrule
     \end{tabular}
  \end{center}
\end{table}



\section{A MIP heuristic for \otsp}
In this section, a MIP heuristic to be solved by off-the-shelf solvers is introduced. The proposed method is solver-independent, but heavily relies on the ability of any given solver to solve the sub-MILPs generated. We first introduce a general \otsp formulation, to then derive the sub-MILPs' formulation. The main idea is fixing variables based on problem-specific dual indicators as proposed in \cite{Fuller2012}, \cite{Ruiz2011} and \cite{Liu2012b}. Row reduction is implicit in sub-MILPs to obtain a tight formulation, eliminating inactive Big-M constraints.

\label{sec:math_framework}
\subsection{Formulation of the OTSP}

The starting point for the development of a heuristic is a generalized \otsp formulation by Fisher et al. \cite{Fisher2008}. {N-1 security is an important criterion for TSOs}. In case of a line or generator outage, the system must stay operational. Let $\C$ denote the set of contingencies, with $a_{ec}\in\{0,1\}$ and $a_{gc}\in\{0,1\}$ denoting the operational status of lines and generators. {For N-1 security, we have that 
\begin{equation}
    \sum_\G a_{gc} + \sum_\E a_{ec} = \lvert \E \rvert + \lvert \G \rvert - 1,~\forall~c\in\C.
\end{equation}}
Transmission switching is NP-hard on its own. Considering multiple scenarios like contingencies results in problems of great complexity. Recent methods focus on decomposing the resulting problem and only considering a small subset of contingencies $\bar{\C}\subseteq\C$, solving small subproblems for every $c\in\bar{\C}$ \cite{Saavedra2020}. With switching as a post-contingency response, these subproblems can become very complex. Any improvements to solve switching problems, hence can help to solve security-constrained problems more efficiently.
A full \otsp formulation for a specific contingency $c\in\mathcal{C}$ or pre-contingency state is presented in \eqref{eq:OriginalTsFormulation}. As we focus on switching problems in this work specifically, we later on assume that $a_{gc}=\bld{1}$ and $a_{ec}=\bld{1}$. We refer to this model as \ModelOne.

\begin{model}[!h]
\caption{OTSP ($\case$) \hfill[MILP]}
\label{Model1}
\begin{subequations}
\label{eq:OriginalTsFormulation}
\begin{IEEEeqnarray}{r'l}   
      \min~z_1 = \sum_{\G} c_{g}(p_g), &
      \text{{s.t.:} } \label{eq:tsObj} \\
      \sum_{\E^{f}_{v}} f_{e} - \sum_{\E^{t}_{v}} f_{e} = \sum_{\G_{v}} p_{g} - d_{v},&\forall\text{ } v \in \V\label{eq:tsMC}\\
      b_{e}(\theta_{v^{f}_{e}} - \theta_{v^{t}_{e}}) + (1-x_{e}a_{ec})M_{e} \geq f_{e},&\forall\text{ } e \in \E\label{eq:tsVoltage1}\\
      b_{e}(\theta_{v^{f}_{e}} - \theta_{v^{t}_{e}}) \leq f_{e} + (1-x_{e}a_{ec})M_{e},&\forall\text{ } e \in \E\label{eq:tsVoltage2}\\
      -\overline{f}x_{e}a_{ec} \leq f_{e} \leq \overline{f}x_{e}a_{ec},&\forall\text{ } e \in \E\label{eq:tsFlowLimit}\\
       \underline{p}_{g}a_{gc} \leq p_{g} \leq \overline{p}_{g}a_{gc},&\forall\text{ } g \in \G\label{eq:ts_generation}\\
      \underline{\theta} \leq \theta_{v} \leq \overline{\theta},&\forall\text{ } v \in \V\label{eq:ts_theta}\\
      x_{e}a_{ec} \in \{0,1\},&\forall\text{ } e \in \E\label{eq:tsGamma}
   \end{IEEEeqnarray}
\end{subequations}
\end{model}

The objective \eqref{eq:tsObj} minimizes the costs of power generation.
The cost function $c_{g}(p_{g})$ is typically a linear or quadratic function. As this model is based on the linear DC power flow approximation (DCOPF), it is treated as linear in the following.
Constraints \eqref{eq:tsMC} are the nodal power balance equations. 
Binary variable $x_{e}$ represents the operational status of line $e$, the switching control.
If $x_{e} = 1$, line $e$ is operational.
If $x_{e} = 0$, it is switched out of service.
If $x_{e} = 0 \rightarrow f_{e} = 0$, due to constraints \eqref{eq:tsFlowLimit}.
Constraints \eqref{eq:tsVoltage1} and \eqref{eq:tsVoltage2} denote the line flow constraints.
Here, $x_{e}$ and $M_{e}$
ensure that these constraints only apply if $x_{e} = 1$.
If $x_{e} = 0$, these constraints are not binding. 

In this section, a restricted OTSP formulation (\rotsp) with an arbitrary amount of switchable lines is derived. This enabled us to develop a heuristic, which meets the requirements stated in section \ref{sec. requ}. We start our exposition by first considering the case of lowest complexity; static topologies.

\subsubsection{DCOPF on static topologies}

Let $\Ea \subseteq \E$ denote the set of all active lines, i.e. disregarding those that are not operational. Note, that for transmission switching any topology can be represented by a set $\Ea$.
A static topology is equivalent to solving DCOPF on a topology $\Ea$. Static topology means that switching statuses cannot be changed by the model, certain lines might be switched out of operation, however, i.e. $e\notin\Ea$ ($x_{e} = 0$ in \ModelOne). 

\begin{model}[!h]
\caption{\dcopf($\Ea$) \hfill [LP]}
\label{Model2}
\begin{subequations} 
\label{eq:OriginalDcopfFormulation}
\begin{IEEEeqnarray}{r'l}
    \min~z_2 = \sum_{\G} c_{g}(p_g), 
     &  \text{{s.t.:} } \label{eq:dcopf_obj} \\
     \sum_{\E^{f}_{v}\cap\Ea} f_{e} - \sum_{\E^{t}_{v}\cap\Ea} f_{e} = \sum_{\G_{v}} p_{g} - d_{v},&\forall\text{ } v \in \V~[\pi]\\
    b_{e}(\theta_{v^{f}_{e}} - \theta_{v^{t}_{e}}) = f_{e},&\forall\text{ } e \in \Ea \label{eq:dcopf_voltageLaw}\\
    -\overline{f}_{e} \leq f_{e} \leq \overline{f}_{e},&\forall\text{ } e \in \Ea\label{eq:dcopf_powerFlow}\\
         \underline{p}_{g} \leq p_{g} \leq \overline{p}_{g} ,&\forall\text{ } g \in \G\\
    \underline{\theta} \leq \theta_{v} \leq \overline{\theta},&\forall\text{ } v \in \V
\end{IEEEeqnarray}
\end{subequations}
\end{model}

Based on this dependency, we define indicator function $\indf:~\E\rightarrow\{0,1\}$. Using $\indf$ every set $\Ea$ can be converted to a corresponding vector $x$, with $\indf(e)=x_{e}$. Note, that this property likewise facilitates constructing a corresponding set $\Ea$ based on a vector $x\in\{0,1\}^{\lvert\E\rvert}$.
A static topology DCOPF formulation can be derived using an arbitrary vector of binary variables $x$. Replacing these variables in \ModelOne with the values in $x$, yields \ModelTwo.

\begin{remark}[\textbf{Feasible Solution Construction from \ModelTwo}]
If a solution $(p^{*}, f^{*}, \theta^{*})$  exists for the \ModelTwo, then $(p^{*}, f', \theta^{*}, x')$,  is a feasible solution for the \ModelOne, where  $f'$, and $x'$ are constructed as follows:

\vspace{-0.2cm}
\begin{IEEEeqnarray}{ll}
   x'_{e} = \indf(e)~\text{and}~
   f'_e = \begin{cases}f^*_e,~e\in \Ea\\0,~e\notin \Ea\end{cases}\label{eq:construction1}
\end{IEEEeqnarray}
\end{remark}

\subsection{Arbitrarily restrictable \otsp formulation}
\label{sec:dcopfOTS}

\begin{figure*}[h]
  \begin{center}
    \begin{subfigure}[b]{0.49\linewidth}
         \centering
         \includegraphics[width=\linewidth]{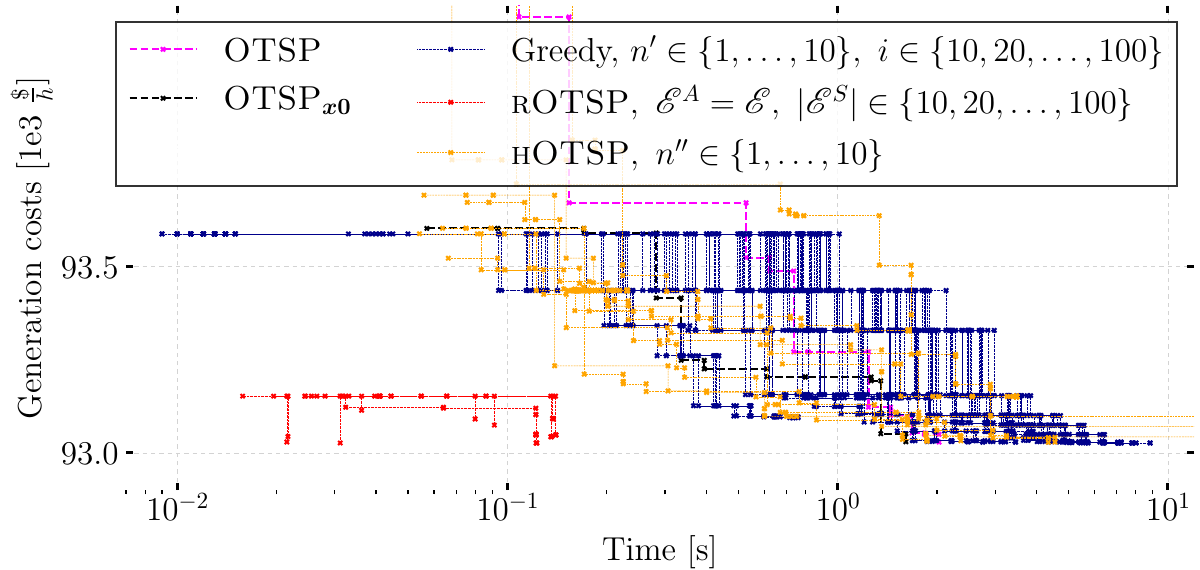}
         \caption{118-bus}
         \label{fig:compHeur118}
    \end{subfigure}
    \begin{subfigure}[b]{0.49\linewidth}
         \centering
         \includegraphics[width=\linewidth]{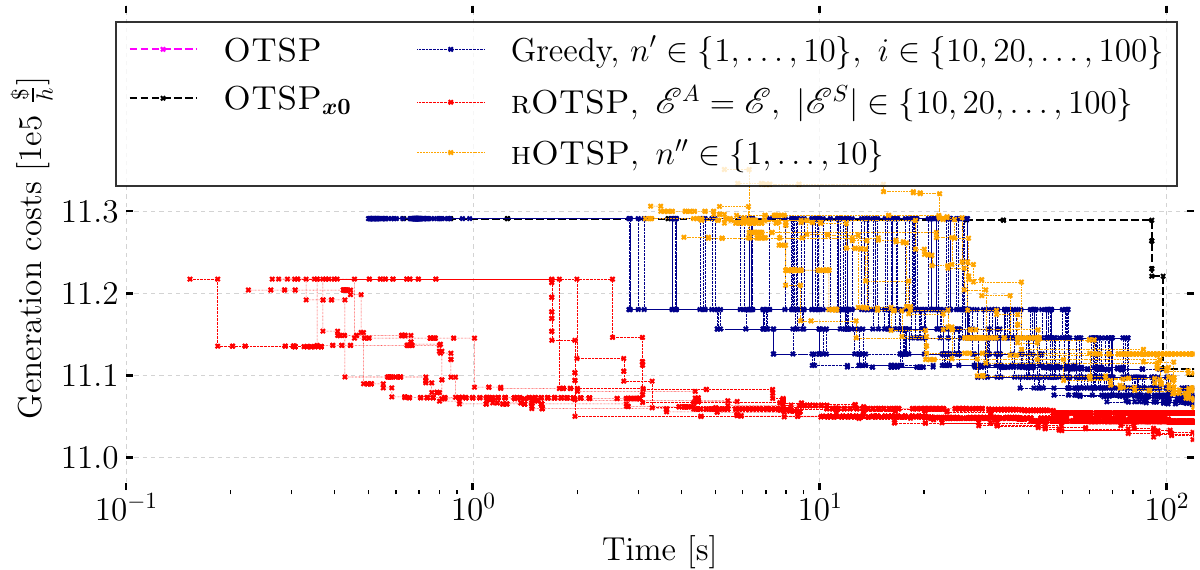}
         \caption{1354-bus}
         \label{fig:compHeur1354}
    \end{subfigure}
    \caption{This plot shows a comparison of different switching heuristics and stock solver performance on the full OTSP, with and without mipstart ($x_0$). To make the comparison as fair as possible, we perform a sweep through parameter values for each heuristic. Whilst we solved \textsc{hOTSP} and \textsc{rOTSP} for 10 different values of $n''$ and $n'$ each (displayed in the legend), we iterated over all $\{1,\dots,10\}\times\{10,20,\dots,100\}$ tuples for the greedy search. This plot clearly shows that \textsc{rOTSP} outperforms both other heuristics by significant margins. All tests were performed using Gurobi to solve the respective LPs and MIPs.}
    \label{fig:compHeur}
  \end{center}
\end{figure*}

We derive an arbitrarily restrictable \otsp formulation (\ModelThree). The main idea is to solve a restricted version \ModelOne that has an easily controllable number of binary variables.  The \ModelThree  is equal to \ModelTwo, if fully restricted to a specific $\Ea$ (no binary variables). Without any restrictions, \ModelThree is equal to \ModelOne.
To achieve the desired behaviour, we must introduce a second set $\Es$ of switchable transmission lines. If line $e\in\Es$, then line $e$ is available for switching. In general,  not all lines are subject to switching. We want to be able to utilize knowledge of any incumbent topology $\Ea$. To derive a model, we use $\ModelTwo$ as a baseline. If a line $e$ should now be subject to switching, a variable $x_e$ and two constraints of type \eqref{eq:tsVoltage1} and \eqref{eq:tsVoltage2} must be attached. If $e\in\Ea$, we must ensure that a constraint \eqref{eq:dcopf_voltageLaw} is not additionally attached to the model. Applying this to all lines in a set $\Es$ yields \ModelThree:
   
\begin{model}[!h]
\caption{Restrictable OTSP: \textsc{rOTSP} ($\Ea$, $\Es$) \hfill [MILP]}
\label{Model4}
\begin{subequations}
\begin{IEEEeqnarray}{r'l}    
      \min z_3 = \sum_{\G} c_{g}(p_g), & \text{{s.t.:}} \\
           \sum_{\E^{f}_{v}\cap\Ea} f_{e} - \sum_{\E^{t}_{v}\cap\Ea} f_{e} = \sum_{\G_{v}} p_{g} - d_{v},&\forall\text{ } v \in \V\\
      b_{e}(\theta_{v^{f}_{e}} - \theta_{v^{t}_{e}}) = f_{e},&\forall\text{ } e \in\Ea\backslash\Es\label{eq:decomposition_voltageLaw1}\\
      b_{e}(\theta_{v^{f}_{e}} - \theta_{v^{t}_{e}}) + (1-x_{e})M_{e} \geq f_{e},&\forall\text{ } e \in\Es\label{eq:decomposition_voltageLaw2}\\
      b_{e}(\theta_{v^{f}_{e}} - \theta_{v^{t}_{e}}) \leq f_{e} + (1-x_{e})M_{e},&\forall\text{ } e \in \Es\label{eq:decomposition_voltageLaw3}\\
      -\overline{f}x_{e} \leq f_{e} \leq \overline{f}x_{e},&\forall\text{ } e \in \Es \quad \label{eq:decomposition_powerFlow1}\\
      -\overline{f} \leq f_{e} \leq \overline{f},&\forall\text{ } e \in \Ea\backslash\Es\label{eq:decomposition_powerFlow2}\\
    \underline{p}_g \leq p_{g} \leq \overline{p}_g,&\forall\text{ } g \in \G\\
    \underline{\theta} \leq \theta_{v} \leq \overline{\theta},&\forall\text{ } v \in \V\\
      x_{e} \in \{0,1\},&\forall\text{ } e \in \Es\label{eq:decomposition_gamma}
   \end{IEEEeqnarray}
\end{subequations}
\end{model}

Constraints \eqref{eq:decomposition_voltageLaw1} and \eqref{eq:decomposition_powerFlow2} are inferred from \ModelTwo, whereas \eqref{eq:decomposition_voltageLaw2}, \eqref{eq:decomposition_voltageLaw3} and \eqref{eq:decomposition_powerFlow1} are used in \ModelOne in the same manner.
The resulting formulation meets all requirements defined in section \ref{sec. requ}. We can start searching for improved solutions by taking advantage of arbitrary incumbent topologies $\Ea$, whilst keeping the constraint matrix compact as constraints and variables are not created for line $e$, iff $e\notin\Ea\cup e\notin\Es$. Furthermore, switching decisions can be changed for line $e$, iff $e\in\Es$. Lastly, the formulation allows us to form \ModelOne whereby a global optimum can be found by exact methods in a finite number of steps. It can be noted, that $\Es\cap\Ea$ is regularly non-empty.
Several important properties of \ModelThree must be stated, as these are the foundation of our algorithm.
\begin{remark} \label{obser1}
\ModelThree equals \ModelOne, if $\Es = \E$.
\ModelThree equals \ModelTwo, if $\Es = \emptyset$.
\end{remark} 
 
 Note, that \ModelThree is a MILP.  Thus, as we increase the size of the set of switchable lines, $n = \rvert\Es\lvert$, the complexity of the model increases. However, it was found \cite{Barrows2012} that only a small set of lines can induce most (or all) of the cost reduction. Similar to the previous \ModelTwo model, we can build feasible solutions for \ModelOne.
 
\begin{remark}[\textbf{Feasible Solution Construction from \ModelThree}]
\label{remark2}
If a solution $(p^{*}, f^{*}, \theta^{*}, x^{*})$ exists for the \ModelThree, then $(p^{*}, f'', \theta^{*}, x'')$,  is a feasible solution for the \ModelOne, where  $f''$, and $x''$ are computed as follows:
\begin{equation}
\footnotesize
  \label{eq:construction2}
   x''_{e} =
  \begin{cases}
      x_{e}^{*}, & e \in \Es \\
      \indf(e), & e \notin \Es \\ 
  \end{cases},~
  f''_{e} =
  \begin{cases}
      f_{e}^{*}, & e \in \Es \\ 
      \begin{cases}f^*_e, e\in\Ea\\0, e\notin\Ea\end{cases}, & e \notin \Es. 
  \end{cases}
\end{equation}

\end{remark}

Two  important properties related to the objective value $z_3$ of \ModelThree, namely, upper bound and monotonicity, can be inferred from the respective model and Observation~\ref{obser1}. Both properties, as previously mentioned, are of great importance to the data exchange at runtime and, hence, facilitate the architecture as well as resulting speedups. We state these properties as remarks, as they must be mentioned for later reference but should be obvious without a proof.

\begin{remark}[\textbf{Upper Bound}] \label{rem.UB}
     Given a solution from \ModelThree, it holds that $z_1^* \leq z_3^*$ for any set $\Es\subseteq\E$.
 \end{remark}
 
\begin{remark}[\textbf{Monotonicity Property}]
    Let $\mathcal{P}_{1}, \mathcal{P}_{2}$ be two instances of the \ModelThree, constructed using the sets $\Ea$ 
    and $\E_{\mathcal{P}_1}^{\text{S}}$ and $\E_{\mathcal{P}_2}^{\text{S}}$, respectively, for each problem instance.
    If $\E^{S}_{\mathcal{P}_1} \subseteq \E^{S}_{\mathcal{P}_2}$, then $z^*_{3}(\Ea,\E_{\mathcal{P}_2}^{\text{S}}) \leq z^*_{3}(\Ea,\E_{\mathcal{P}_1}^{\text{S}})$.
\end{remark}
 
Given the aforementioned properties, the optimal objective values of the constructed problems are progressively closer to the optimal objective value of the original problem \ModelOne, with an increasing $n$ (number of switchable lines). However, our heuristic relies on the fact that often only a couple of lines must be switchable (included into $\Es$) to achieve close-to-optimal solutions.
The literature on transmission switching heuristics develops methods like iterative schemes and MIP heuristics as well as switching indicators to identify promising switching candidates. In the next subsection, we compare different heuristic methods against solving the full problem for different test cases.
We then compare the effectiveness of line switching indicators using our proposed MIP heuristic to identify a superior indicator, confirming results in \cite{Ruiz2011}.

\subsection{Comparison of switching heuristics}

\begin{figure*}
     \centering
     \begin{subfigure}[b]{0.245\linewidth}
         \centering
         \includegraphics[width=\linewidth]{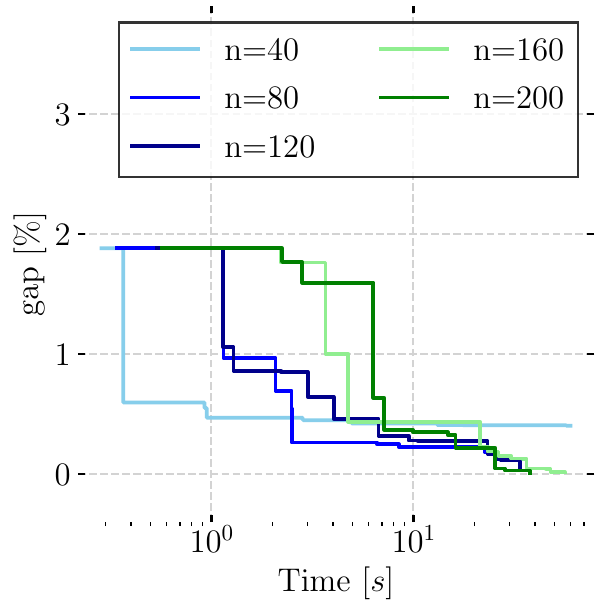}
         \caption{Line profits}
         \label{fig:sc1}
     \end{subfigure}
     \begin{subfigure}[b]{0.245\linewidth}
         \centering
         \includegraphics[width=\linewidth]{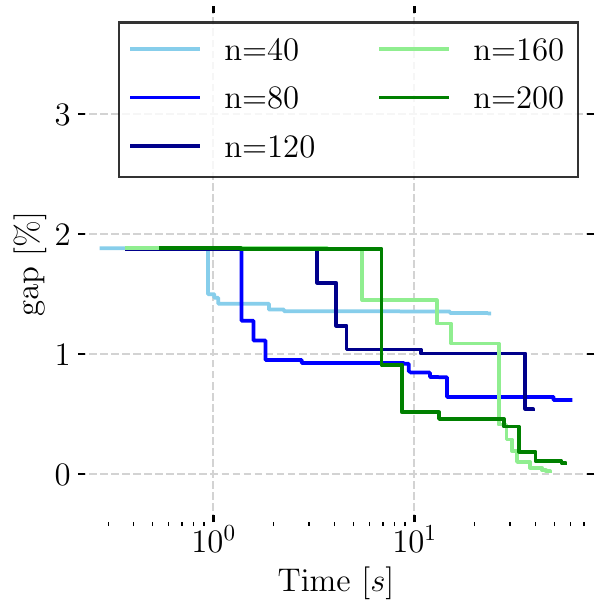}
         \caption{Price difference}
         \label{fig:sc2}
     \end{subfigure}
     \begin{subfigure}[b]{0.245\linewidth}
         \centering
         \includegraphics[width=\linewidth]{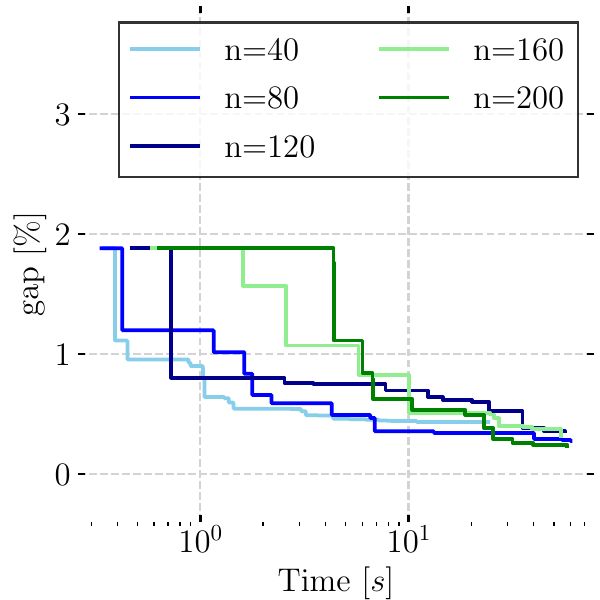}
         \caption{Total cost derivative}
         \label{fig:sc3}
     \end{subfigure}
     \begin{subfigure}[b]{0.245\linewidth}
         \centering
         \includegraphics[width=\linewidth]{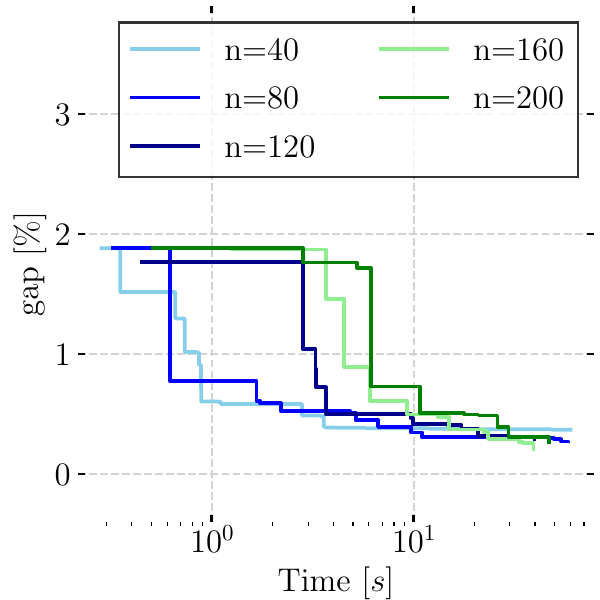}
         \caption{PTDF-weighted}
         \label{fig:sc4}
     \end{subfigure}
        \caption{This figure depicts the solution process of solving \ModelThree on the 1354\_pegase case with different numbers of switchable lines $n = \lvert\Es\rvert$, using Gurobi.  Additionally, we compare solver performance, using different switching criteria  to select lines $e\in\Es$. Every marker represents an integer-feasible solution being found.}
        \label{fig:compHeur}
\end{figure*}

We now compare two heuristics to \ModelThree without any parallel implementation. In this case, \ModelThree is similar to previously developed prescreening methods \cite{Liu2012b}. 
One of the most common heuristics is an iterative greedy search that identifies a single line to be removed based on an indicator, as was also brought forward in \cite{Fuller2012}. This heuristic specifically has two main control parameters: $n'$ denotes the maximal number of lines to be removed (outer iterations), while $i$ represents the number of lines to be evaluated based on a defined criterion before selecting the next line for removal (inner iterations). The higher these numbers the more ($n' \times i$) iterations are performed.
\newline
A common MIP heuristic that is used not only in \cite{Fuller2012} as a benchmark, but also in \cite{Hedman2009} to deal with the OTSP's complexity, is the addition of an additional constraint to produce an altered version of \ModelOne; presented as \ModelFour.
\begin{model}[!h]
\caption{Heuristic OTSP: \textsc{hOTSP} ($\case$) \hfill [MILP]}
\label{Model4}
\begin{subequations}
\begin{IEEEeqnarray}{r'l}    
      \eqref{eq:tsObj} - \eqref{eq:tsGamma}~&\\
      \sum_{e\in\E}(1-x_{e}) = n'' \label{eq:cut}
   \end{IEEEeqnarray}
\end{subequations}
\end{model}
Here, the parameter $n''$ denotes the number of lines to be switched out of operation simultaneously.
These heuristics are now compared against \ModelThree with similar parameters. The switching criterion used for \ModelThree and the iterative greedy search is the line profit criterion to be introduced in the next section. This was found superior in previous research already \cite{Ruiz2011}, \cite{Fuller2012}.

The comparison in Figure \ref{fig:compHeur1354} reveals significant differences between the tested methods. First, the MIP heuristics are easier to tune to a specific problem as they have fewer parameters. Their downside is that their performance is largely dependent on the MIP solver used, while the greedy search just requires an LP solver. Cuts \eqref{eq:cut} result in models that are still very complex; as $\frac{\lvert\E\rvert!}{n!(\lvert\E\rvert-n)!}$ ways exist to remove $n$ lines from the system. \ModelThree on the contrary just results in $2^n$ options. Worse, constraints \eqref{eq:cut}, while restricting the search space, resulting in more infeasible vectors $x$ when compared to the original problem. This increases the complexity of the problem and effectiveness of internal solver heuristics as they are more likely to produce solutions violating additional constraint(s). Figure \ref{fig:compHeur118} illustrates that such constraints can even increase computation times over solving the full formulation. It must be noted, however, that most solutions are produced by internal heuristics of a MIP solver. MIP heuristics take full advantage of MIP solvers' capabilities. \ModelThree combines the advantages of switching indicators and MIP heuristics. Even though the global solution is cut off, we can achieve very good results on the 118-bus and 1354-bus cases by switching at most 10 or 30 lines respectively. All other methods fall short by a significant margin in computational time.

\subsection{Selecting candidate lines for transmission switching}
\label{sec:cts}
To use \ModelThree, we must construct a set $\Es$. We want to include only lines for switching that decrease operational costs. In the following, we require power transfer distribution (PTDF) and line outage distribution factors (LODF). The cardinality of $\Es$ is a direct driver of the computational complexity of the resulting \ModelThree.
To identify candidate lines, four switching criteria have been proposed in the literature:
The \textit{line profits switching criterion} (\lpsc) as proposed in \cite{Fuller2012} and \cite{Ruiz2011} is derived using KKT-conditions and expressed as:
\begin{equation}
\alpha_e = f^{*}_e(\pi^*_{v^f_e} - \pi^*_{v^t_e}).\label{eq:lpsc}
\end{equation}
Additionally, a \textit{price difference}, \textit{total cost derivative} and \textit{PTDF-weighted cost derivative switching criterion} were proposed in \cite{Ruiz2011}.
Removing a candidate line based on these criteria does not guarantee improvements due to the problem being discrete in nature. Removing a line might render the problem infeasible.
Deriving indicators for all lines facilitates to create a sorted list $\lp$, in ascending order, by the lines' respective switching criterion (highest reduction first).
All criteria can be computed based on \ModelTwo and its dual variables.
Based on $\lp$, we create a subset of switchable lines $\Es$, by selecting the $n$ first lines in $\lp$.
To illustrate the effect of different heuristics and size of  the switching lines set,  we tested all switching criteria using \ModelThree with different values for $n$. The results of the case 1354\_pegase are shown in Figure~\ref{fig:compHeur}.
Results indicate, that the \lpsc performs best. Especially for smaller sets of switchable lines, the improvements to be achieved are higher than with other criteria. The total cost derivative switching criterion also yields similar results, but only for larger sets $\Es$ ($n=160,~n=200$). The total cost derivative and PTDF-weighted criteria in Figures \ref{fig:sc3} and \ref{fig:sc4} do not produce comparable improvements and are more expensive to compute, as they require the PTDF and LODF respectively.

\section{Algorithmic Architecture and Parallel Implementation}
\label{sec:algorithm_arqu}

\subsection{Parallel Framework}
\label{sec:parallelFramework}

As a framework for parallel high-performance computing, the open-source implementation message passing interface chameleon (MPICH) is used.
This tool facilitates defining independent processes (also called ranks), implementing communication, and allocating resources.
In our design, multiple ranks can be defined as 0 and $\{1,\dots,j\, \ldots, |\J|\}$. We reserve rank 0 for the principal worker, the main solution process. Information is exchanged between rank 0 and the others asynchronously. Whereas rank 0 receives new solutions from a rank $j$, the global incumbent and information on the solution process are sent from rank 0 to all $j\in\J$ continuously.
To implement advanced solver behavior (including communication), Gurobi's user callbacks are utilized. Callbacks facilitate online communication between external software and the MIP solver, i.e., sending and retrieving data during runtime. Here, external software might also mean an independent instance of the same MIP solver. Calling user functions is possible during different times of the iterative solution process, after exploring a MIP node or finding a new solution, for example. While parallel implementations might require a process that assigns workloads \cite{Namazifar2008}, callbacks and functions implemented in MPI allow for completely asynchronous behavior, with optimizers running on all ranks.

\begin{figure}[h]
  \begin{center}
    \includegraphics[width=.98\linewidth]{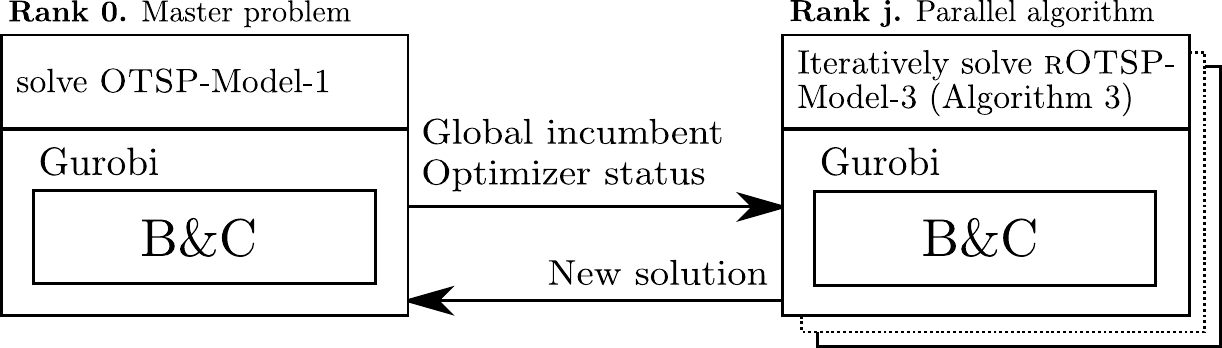}
    \caption{Parallel coordination}
    \label{fig:simpleFlowChart}
  \end{center}
\end{figure}

Figure \ref{fig:simpleFlowChart} represents a simplified version of the implemented parallel coordination and data exchange.
The coordination can be summarized as follows: while solving \ModelOne, we solve instances of \ModelThree in parallel. Since \ModelThree can be constructed with a reduced number of integer variables, feasible solutions can be obtained quicker. If a solution is obtained on any rank, we convert it to a solution of \ModelOne (see the remark \ref{remark2}) and send it to rank 0. It may occur that the solver instance on rank 0 attained an upper bound that is better than any solution expected to be produced by \ModelThree on a rank j. In this case, we send the superior solution (global incumbent) to rank j, build a new instance of \ModelThree and continue the process.
The parallel algorithm terminates after reaching the time limit or obtaining an optimal solution at rank 0. Then again, a signal is sent to all workers to stop optimization.
In the next subsection, we present the parallel algorithm in detail.

\subsection{Parallel Algorithm (Rank $j$)}
\label{sec:heuristics}
The main goal of our parallel algorithm is to provide strong solutions for \ModelOne as fast as possible. 
This (feasible) solution might update the upper bound of the OTSP in rank 0. To keep algorithms concise, several technical remarks must be made on how communication is implemented. 
When sending data using MPI, tags are set to facilitate the identification of the data. In the proposed architecture, 3 pieces of information are exchanged and must be identified: The optimization status information (termination) at rank 0 will be referred to in the following as \stat, solutions produced at rank j by {\sol}, and global incumbents by \inc. A pseudo-code call Recv(\textit{src}, \textit{tag}) entails probing for data of type \textit{tag} sent from source \textit{src}, to then receive it if it is available. Similarly, a call Send(\textit{data}, \textit{dst}, \textit{tag}) means that \textit{data} is sent to the destination \textit{dst}. On several occasions, we also wait for requests to be completed. Furthermore, let $\case$ denote all parameters related to a test case.
Our algorithm is executed on ranks j and consists of three major components: (i) initialization, (ii) prioritization, and (iii) solving restricted \rotsp instances.

\subsubsection{Initialization}

As described, we rely on a feasible set $\Ea$. The feasibility of such a set can be validated by solving \ModelTwo.
From the solution obtained, we can build a feasible solution for \ModelOne, as described in \eqref{eq:construction1} and derive dual values.
The procedure is summarized in \mbox{\AlgOne}.
Note that $x'$ is used as \textit{mipstart} (initialization of binary variables) in subsequent steps (\AlgThree). 
\begin{algorithm}
  \SetKwInOut{Input}{Input}\SetKwInOut{Output}{Output}\SetKwInOut{Requires}{Requires}
  \Input{$\case$, $\Ea$\\}
   $(p^{*}, f^{*}, \theta^{*}, \pi^*)$  $\leftarrow$ \textbf{solve } \ModelTwo($\case$,~$\Ea$)\\
  $(f', x')$ = Generate feasible solution (Rem. \eqref{eq:construction1})\\
   \Output{$(p^{*}, f', \theta^{*}, x',\pi^*)$}
  \caption{Initialization Procedure}
  \label{alg:heuristic1}
\end{algorithm}
\vspace{1em}
%

\subsubsection{Prioritization of line selection for switching}

Based on the results presented in section \ref{sec:cts}, the \lpsc criterion is chosen for line prioritization. To compute priority value $\alpha_{e}$ based on \eqref{sec:cts}, LMPs $\pi^*$ and power flow values $f^*$ must be obtained by solving \ModelTwo.
\AlgTwo describes how we build a priority list $\lp$, based on the \lpsc derived from \ModelTwo.

\begin{algorithm}
  \SetKwInOut{Input}{Input}\SetKwInOut{Output}{Output}\SetKwInOut{Requires}{Requires}
  \Input{$\case$, $\Ea$, $n$}
    \tcc{Note, that $\pi^{*}$ and $f^{*}$ only exist if \ModelTwo($\case,~\Ea$) has a solution.}
    $(p^{*}, f', \theta^{*}, x',\pi^*) \leftarrow \AlgOne(\case, \Ea$)\\
   $\alpha = 0^{\lvert\E\rvert}$, $\Es=\emptyset$\\
   \For{$e \in \E$}{
   \tcc{Any (switching) criterion can be used.}
        $\alpha_{e} = f'_e(\pi^*_{v^{f}_{e}} - \pi^*_{v^{t}_{e}})$
    }
   $\lp$ $=$ sort($\E$, by $\alpha$, ascending) \\
   $\Es \leftarrow \lp_{1,\dots,n}$\\
   \Output{$\Es$}
   \caption{Line Prioritization}
  \label{alg:heuristic2}
\end{algorithm}
\vspace{1em}

\subsubsection{Solving restricted OTSP instances} 

\AlgThree shows how the information of each rank is exchanged to speed up both B\&B algorithms. We use $\textbf{R0}$ and $\textbf{Rj}$ to refer to rank 0 and rank j, respectively.

\AlgThree is running as long as the master problem \ModelOne is running. A binary variable \textit{root\_process\_terminated} indicates whether solving the master problem is still ongoing. Its state, along with the global incumbent, is continuously sent from rank 0. The solution process on rank 0 is terminated upon reaching a time limit or obtaining an optimal solution.

The input of \AlgThree is the electric grid data and parameters $n\in\mathbb{N}^{\text{+}}$ (number of switchable lines to be added to the set $\Es$) and $\Delta{n}\in\mathbb{N}^{\text{+}}$ (number of new lines to added to the set $\Es$ after every iteration). The parameter  \textit{update\_time} (time window to retrieve the global incumbent from rank 0 - fixed to 10s in our case) and \textit{reset\_time} (time window to check if no better solution than rank 0 can be found in rank 1- fixed to 20 s in our case) must be set as well.

\AlgThree starts with the initialization of sets of switchable lines $\Es = \emptyset$ and active lines $\Ea = E$. Based on $\Ea$ a \textit{mipstart} is generated for \ModelThree using \AlgOne and the priority list $\lp$ using \AlgTwo. Then, iteratively, we solve \ModelThree and pass its solutions to the main B\&C instance at rank 0. 


\begin{algorithm}
  \SetKwInOut{Input}{Input}\SetKwInOut{Output}{Output}\SetKwInOut{Initialize}{Initialize}
  \Input{$\mathbb{c}$, $n$, $\Delta{n}$  \\
  update\_time, reset\_time}  
  \Initialize{ $\Ea = \E$\\
  $(p^{*}, f', \theta^{*}, x',\pi^*) \leftarrow \AlgOne(\case, \Ea)$\\
  Send\big($(p, f', \theta, x')$, \textbf{R0}, \sol\big)\\
  \textit{mipstart} $= x'$\\
  $\Es \xleftarrow{}$ \AlgTwo($\case,\Ea,n$)}
  \While{$\neg$(root\_process\_terminated)}{
    \tcc{\textbf{(a)} Retrieve global incumbent from rank 0, if available}
    $(ub^{\textbf{R0}}, p^{\textbf{R0}}, f^{\textbf{R0}}, \theta^{\textbf{R0}}, x^{\textbf{R0}})$ = Recv(\textbf{R0}, \inc)\\
    $\Ea$ = $\{e\in \E\mid x_{e}^{\textbf{R0}} = 1\}$\\
    \textit{mipstart} $=$ $(\indf(e_{1}),\dots,\indf(e_{\lvert\Es\rvert}))$  \\
    \tcc{\textbf{(b)} Construct \ModelThree with current $\Ea$ and $\Es$, and call B\&C optimization.}  
      \textbf{solve }\ModelThree($\case,\Ea$,$\Es$, \textit{mipstart})\\
      \While{solving \ModelThree}{
        \tcc{\textbf{(c)} Update global incumbent from rank 0 every \textit{update\_time}. Get information about if rank 0 returned.}
        \If{update\_time\_passed}{
        $(ub^{\textbf{R0}}, p^{\textbf{R0}}, f^{\textbf{R0}}, \theta^{\textbf{R0}}, x^{\textbf{R0}})$ = Recv(\textbf{R0}, \inc)\\
        root\_process\_terminated = Recv(\textbf{R0}, \stat)\\
        }
        \tcc{\textbf{(d)} Reset \ModelThree B\&C if $lb^{\textbf{Rj}} \geq ub^{\textbf{R0}}$ }
        \If{reset\_time\_passed}{
            \If{Condition in \textbf{(d)} met}{
                Stop B\&C of \ModelThree\\ 
            }
        }
        \tcc{\textbf{(e)} Send feasible solution to rank 0 when new MIP solution, $x^{\textbf{Rj}}$, is found in \ModelThree with objective value $z^\textbf{Rj}$ better than $ub^{\textbf{R0}}$}
         \If{New sol. found $\land$ $z^\textbf{Rj} < ub^{\textbf{R0}}$}{
                  $(p, f'', \theta, x'')$ $\xleftarrow{}$ Generate feasible solution from \eqref{eq:construction2}\\
                Send\big($(p, f'', \theta, x'')$, \textbf{R0}, \sol\big)\\
         }
      }
    \tcc{\textbf{(f)} Update $\Es$}
    $n = \min\{n + \Delta{n}, \lvert\E\rvert \}$ \\
    $\Ea$ = $\{e\in \E\mid x_{e}^{\textbf{R0}} = 1\}$\\
    $\Es \leftarrow{}$ \AlgTwo($\case,\Ea,n$) \\
  }
\caption{Priority-Guided Heuristic}
\label{alg:heuristic3}
\end{algorithm}
\vspace{1em}

While solving \ModelThree, six main tasks are performed. Three out of six are within the scope of the implemented callback function. They are described below. 

\paragraph{Retrieving global incumbent from rank 0, if available} Rank 0 continuously sends the incumbent to \ModelOne. If an incumbent was sent, it is received by rank j and used to update \textit{mipstart}, $\Ea$ and information on the global \ub at rank 0 saved by rank j, $ub^{\textbf{R0}}$. 

\paragraph{Reconstruction of \ModelThree with updated set information for $\Ea$, $\Es$ and \textit{mipstart}}. First, \ModelThree is loaded into a MIP solver (Gurobi in our case). On loading the model, a callback function is attached. This function performs 3 main behaviors, dependent on the current solver state.

\paragraph{Update the global incumbent retrieved from rank 0 every \textit{update\_time}} In regular intervals, it is checked whether a new incumbent was sent from \textbf{R0}. If so, \textbf{Rj} receives it and updates its global incumbent information. This information is vital to control the behavior of worker ranks. Upper bound $ub^{\textbf{R0}}$ is required, for example, to determine if better solutions can still be found on rank j with the current instance of \ModelThree, step (d).

\paragraph{Reset the B\&C algorithm at rank j}\label{sec:reset} We use 2 criteria to abort optimization of our MIP heuristic. These conditions are evaluated continuously after an interval of length \textit{reset\_time} has passed. First, $lb^{\textbf{Rj}} \geq ub^{\textbf{R0}}$ means that no new incumbents can be produced by our MIP heuristic. If that condition is met or no improvements were found during the last 20 s, we terminate optimization and go to step (f).  

\paragraph{Send feasible solution to rank j}  Whenever a feasible solution is found for \ModelThree during B\&C exploration or internal heuristics, it is converted (according to \eqref{eq:construction2}) and sent to rank 0. Rank 0 then updates its incumbent based on this solution, if $z^\textbf{Rj} < ub^{\textbf{R0}}$. 

\paragraph{Update $\Es$} 
Finally, the ordered list of switching lines $\lp$ is updated. The new set of candidate lines for switching control is updated taking the first $n$ lines. The number of lines $n$ is increased after each iteration by $\Delta{n}$. Eventually, in the worst case, $n$ could be equal to the set of all lines, and recover \ModelOne.

\section{Numerical Experiments}
  \label{sec:numerical_results}

\subsection{Test cases and setup}
Hard- and software-wise, our test system has 16 GB of RAM and is equipped with an i7-7820X, clocked at 4.0 GHz.
As all experiments were run with hyper-threading disabled, a total of 8 logical cores were available.
The following software (packages) were used in our work: Gurobi 9.1.2 \cite{gurobi}, MPICH 4.3.1 \cite{mpich}, Julia 1.6.3 \cite{Julia2017}, JuMP.jl 0.21.2 \cite{DunningHuchetteLubin2017}, Gurobi.jl 0.9.14, MPI.jl 0.19.1 \cite{Byrne2021} and Matplotlib 3.0 \cite{Hunter2007}.

For all experiments, the \verb+pglib-opf+ library was used. We set $\underline{p}=\boldsymbol{0}$.
Software for managing data files and running our experiments was written in  \textit{Julia Language}\footnote{The implementations is accessible here: \url{https://github.com/antonhinneck/PowerGrids.jl}, \url{https://github.com/antonhinneck/ParallelHeuristics.jl}}.

\subsection{Computational performance with consistent parameters}
{\def\arraystretch{1.35}\tabcolsep=5.5pt
\begin{table*}[ht!]
   \renewcommand\thetable{II}
   \caption{Experimental results}
   \label{tab:results}
   \begin{center}
     \begin{tabular}{ |c | c c c | c c c | c c c | c c c c|}
        \hline
        \multicolumn{4}{|c}{} & \multicolumn{3}{c}{\textbf{OTSP, no mipstart}} & \multicolumn{3}{c}{\textbf{$\otsp_{\xn}$, mipstart}} & \multicolumn{4}{c|}{\textbf{P-OTSP}}\\
        \hline
        \# & \textbf{Case} & \textbf{\# buses} & \textbf{\# lines} & \textbf{ct [s]} & \textbf{gap [\%]} & \textbf{$\Delta z$ [\%]} & \textbf{ct [s]} & \textbf{gap [\%]} & \textbf{$\Delta z$ [\%]} & \textbf{ct [s]} & \textbf{gap [\%]} & \textbf{$\Delta z$ [\%]} & $\overline{\Delta z}[\$/h]$\\
        \hline 
        1 & 118\_ieee & 118 & 186 & 0.72 & 0.0$^{\text{+}}$ & 13.491 & 1.53 & 0.0$^{\text{+}}$ & 13.491 & 1.82 & 0.0$^{\text{+}}$ & 13.491 & 0.0\\
        2 & 588\_sdet & 588 & 686 & 900 & 0.064 & 2.14 & 900 & 0.022 & 2.19 & 900 & 0.039 & 2.17 & -36.65\\
        3 & 1354\_pegase & 1354 & 1991 & 900 & 0.04 & 1.938 & 900 & 0.1 & 1.882 & 192.03 & 0.007$^{\text{+}}$ & 1.971 & 364.13\\
        4 & 1888\_rte & 1888 & 2531 & 369 & 0.0092$^{\text{+}}$ & -0.0092 & 0.39 & 0.0$^{\text{+}}$ & 0.0 & 2.03 & 0.0$^{\text{+}}$ & 0.0 & 0.0\\
        6 & 2383\_wp\_k & 2383 & 2896 & 900 & - & - & 900 & 0.18 & 4.05 & 900 & 0.02 & 4.21 & 2732.38\\
        7 & 2736\_sp\_k & 2736 & 3504 & 900 & 0.07 & 6.82 & 900 & 0.65 & 6.21 & 900 & 0.04 & 6.86 & 355.1\\
        8 & 2746\_wop\_k & 2746 & 3514 & 900 & - & - & 900 & 0.19 & 10.24 & 494.5 & 0.003$^{\text{+}}$ & 10.46 & 2094.06\\
        9 & 2869\_pegase & 2869 & 4582 & 900 & - & - & 900 & 0.39 & 1.07 & 900 & 0.18 & 1.27 & 4817.47\\
        10 & 3012wp\_k & 3012 & 3572 & 900 & 0.54 & 8.01 & 900 & 0.44 & 8.11 & 900 & 0.31 & 8.22 & 2870.06\\
        11 & 3120sp\_k & 3120 & 3693 & 900 & 0.7 & 6.52 & 900 & 0.9 & 6.4 & 900 & 0.29 & 6.77 & 5510.58\\
        12 & 3375sp\_k & 3375 & 4161 & 900 & 1.34 & 3.15 & 900 & 1.17 & 3.32 & 900 & 0.86 & 3.62 & 19966.29\\
        13 & 6470\_rte & 6470 & 9005 & 900 & - & - & 900 & 8.83 & 4.39 & 900 & 5.32 & 7.46 & 55207.37\\
        14 & 13659\_pegase & 13659 & 20467 & 900 & - & - & 900 & 0.88 & 0.0 & 900 & 0.42 & 0.46 & 31022.55\\
        \hline 
     \end{tabular}
     \vspace{0.5ex}
     \footnotesize{
     \newline
     $^{\text{+}}$ The solution is classified optimal as \textbf{gap} $\leq0.01$. A timelimit of 900 s was set for all experiments.}
   \end{center}
\end{table*}
}

\begin{figure*}[h]
  \begin{center}
    \includegraphics[width=1\linewidth]{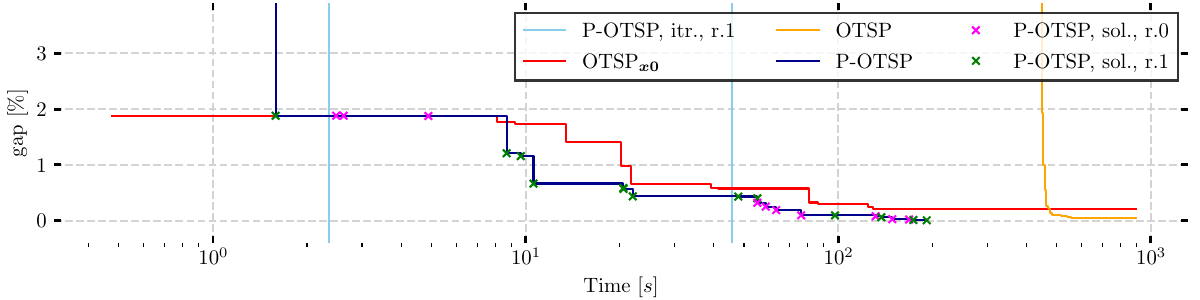}
    \caption{ This plot shows the solution process on the 1354\_pegase test case, utilizing different solution strategies. Both baseline methods (\otsp and $\otsp_{\boldsymbol{x0}}$) do not find an optimal solution within the time limit of 900 s. Supplying a mipstart $x_0$ has a notable effect on solutions that can be obtained, especially early on. Using our parallel implementation (P-OTSP) an optimal solution is found in less than 193 s. The plot shows $\ub$, $\lb$ of the P-OTSP procedure, as well as whether a solution was produced by the solver instance working on the full problem (P-OTSP, sol., r.0) or a MIP heuristic (P-OTSP, sol., r.1). We, furthermore, depict whether a full iteration was completed by \AlgThree (P-OTSP, iter., r.1). At around 45 s, optimization on rank 1 restarts (iteration) due to no solutions being found. We can see that right after restarting optimization from a different $x^{\textbf{R0}}$, new solutions are found quickly.}
    \label{fig:1354_pegase}
  \end{center}
\end{figure*}

We first want to verify, that our method can produce improvements on a multitude of test cases with consistent settings. For the following experiments, we used two ranks, $\J = \{1\}$. Using affinity maps, we pinned rank 0 to cores 0-3 and rank 1 to cores 4-7. As parameters for Algorithm \ref{alg:heuristic3}, $n=40$ and $\Delta n=10$ were chosen.
We set a time limit of 15 minutes. Three exact solution strategies were implemented for solving each test case. These are described in the following:

\begin{itemize}
    \item \textbf{\otsp, no mipstart}. We solve \ModelOne using Gurobi with default settings. No \textit{mipstart} is provided. Gurobi has available all logical cores (8). 
    \item \textbf{$\otsp_{\xn}$, mipstart}. We solve \ModelOne using Gurobi with default settings. An initial solution is generated from \AlgOne and passed to Gurobi as \textit{mipstart}. Gurobi has available all logical cores (8). 
    \item \textbf{P-\otsp}. In this case, we use the parallel architecture introduced in section \ref{sec:algorithm_arqu}.
\end{itemize}
Table \ref{tab:results} shows the results of solving 14 test cases utilizing the three aforementioned strategies. 
The experiment id, the test case's name, and the number of buses and lines are shown in columns 1-4. 
All lines are considered switchable. Thus, the number of lines also denotes the number of integer variables of each problem. The smallest grid has 118 buses and 186 lines, while the largest grid has 13659 buses and 20467 lines. 
Furthermore, solutions with $\gap \leq 0.01 \%$ are classified as {optimal}.
For the non-optimal cases, optimization is terminated upon reaching the time limit. For each of the three solution strategies, we report the computational time ct), relative optimality \gap, and relative cost reduction ($\Delta z$) of the \otsp over a linear DC approximation with all lines active.  In addition, we report the absolute cost reduction of P-OTSP over either \otsp or $\otsp_{\xn}$, depending on which performed better ($\overline{\Delta z}$). 

For\textit{ small test cases} (such as \#1, \#2 and \#4), performance improvements might not be achieved. For the 118-bus case, optimization takes less than 2 seconds until an optimal solution is found. Using optimization packages and additional software for parallel computation, we add supplementary code and therefore run and compile time to the solution process. This becomes less insignificant as the problems are more difficult to solve. In cases as large as the 588\_sdet-bus case, Gurobi still does not benefit from the proposed solution technique. However, our method also does not harm the performance significantly either. This generally starts changing with the 1354-bus case. Case \#4 is large as well, but $\Ea = \E$ is an optimal topology. Our method also has no significant effect, as a certificate of optimality can be attained very quickly. 

For \textit{larger cases} starting with the 1354-bus case, significant improvements in performance or solution quality can be observed. Figure  \ref{fig:1354_pegase} depicts the solution process. While baseline methods yield good solutions with a low final \gap of 0.04\% and 0.07\% respectively, after reaching the time limit of 900 s, P-OTSP finds an optimal solution after 192.03 s. With a final \gap of 0.007\% the difference to both other solution methods seems small but is significant in absolute terms with 364.13 [\$/h]. Figure \ref{fig:1354_pegase} shows that much better solutions are found during the first 100 s. Solutions are found by Gurobi's heuristics on rank 0 and rank 1. Whenever good solutions are supplied, further improvements can quickly be obtained based on them. Throughout the optimization process, Algorithm \ref{alg:heuristic3} performs 3 full iterations, which are displayed in vertical blue lines. This is due to better incumbents being available on rank 0, finding an optimal solution for a particular instance of \ModelThree, or not having found any solution within the last 20 seconds.

\begin{figure}
     \centering
     \begin{subfigure}[b]{\linewidth}
         \centering
         \includegraphics[width=\linewidth]{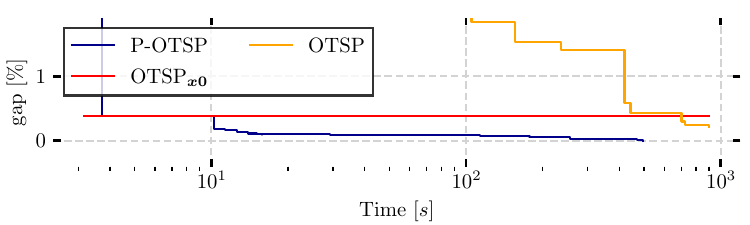}
         \caption{2746wop\_k}
         \label{fig:resp1}
     \end{subfigure}
     \begin{subfigure}[b]{\linewidth}
         \centering
         \includegraphics[width=\linewidth]{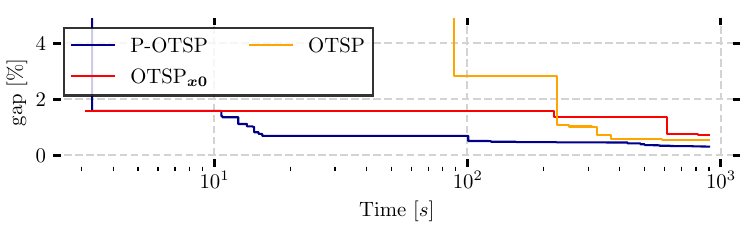}
         \caption{3012wp\_k}
         \label{fig:resp2}
     \end{subfigure}
     \begin{subfigure}[b]{\linewidth}
         \centering
         \includegraphics[width=\linewidth]{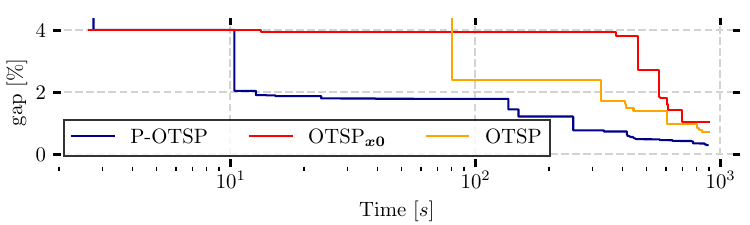}
         \caption{3120sp\_k}
         \label{fig:resp2}
     \end{subfigure}
     \begin{subfigure}[b]{\linewidth}
         \centering
         \includegraphics[width=\linewidth]{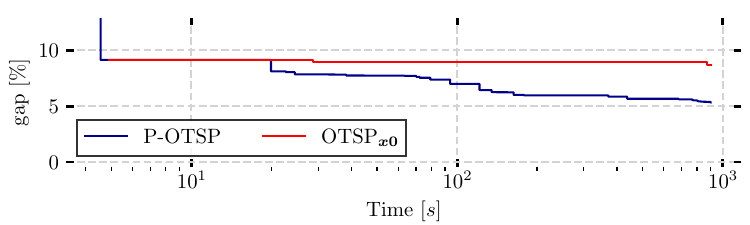}
         \caption{6470\_rte}
         \label{fig:resp3}
     \end{subfigure}
          \begin{subfigure}[b]{\linewidth}
         \centering
         \includegraphics[width=\linewidth]{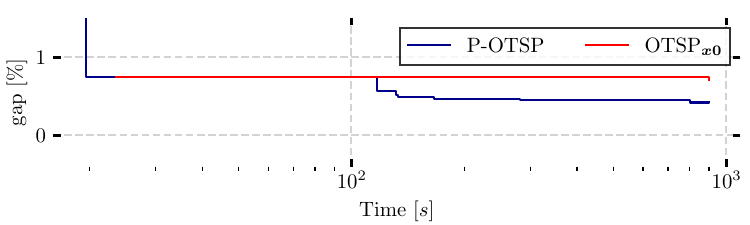}
         \caption{13659\_pegase}
         \label{fig:resp4}
     \end{subfigure}
        \caption{ In this figure, the solution process of 4 additional test cases are illustrated. We plot the \gap for OTSP, $\otsp_{\xn}$ and P-OTSP. Every marker denotes obtaining an integer-feasible solution. For larger problems, the MIP solver is not able to find a feasible to find  solution within the time limit. We can reliably obtain better solutions.}
        \label{fig:selCases}
\end{figure}
Figures \ref{fig:selCases} depict the progress on 4 additional test cases, again for each solution strategy.
Figure \ref{fig:resp1} shows the progress in the largest case an optimal solution can be obtained, within the time limit. The progress of the different methods is similar to that of the 1354-bus case. Without a mipstart solution, it takes much longer until good solutions can be achieved. Our parallel framework not only finds an optimal solution, but extremely good solutions very early on. In this case, a cost reduction of 2094.06 $[\$/h]$ can be achieved.

While solving the 3120-bus case, $\otsp_{\xn}$ is slightly ahead at around 90 s, due to a strong solution supplied by an internal heuristic. We observe, however, that P-OTSP improves on solutions much faster, which results in a final total improvement of 5510.58 $[\$/h]$, over the best competing method. The last displayed test cases 6470\_rte (Fig.\ref{fig:resp3}) and 13659\_pegase (Fig. \ref{fig:resp4}) show very similar solution processes. Without a mipstart, Gurobi cannot find a solution within the time limit. Providing a start solution results in multiple further improvements. P-OTSP outperforms a single solver instance significantly in both cases. Due to the fact that these systems are large and P-OTSP finds significantly better solutions, absolute cost reductions are 55207.37 $[\$/h]$ and 31022.55 $[\$/h]$ over competing methods.
On a general note, significant improvements can be produced utilizing P-OTSP. 

\subsection{Performance using multiple worker ranks}

During the previously described experiments, we only used 1 worker rank, solving a single instance of \ModelThree at a time. Figures \ref{fig:compHeur} show, however, that small sets of switchable lines  ($n=40$) result in fast improvements, whilst larger sets ($n=200$) result in very good solutions. The advantages of a higher degree of parallelization are studied in the 1354\_pegase test case. We now run 1, 2 and 3 worker ranks in parallel to our master problem, each with a different parameter $n_1=40,~n_2=120$ and $n_3=200$. With 2 worker ranks, we pin the root rank to 4 cores and each worker rank to 2, whilst with 3 worker ranks, each rank is pinned to 2 cores. We compare our results to $\otsp_{\xn}$ as well.

\begin{figure}[h]
 \centering
 \includegraphics[width=\linewidth]{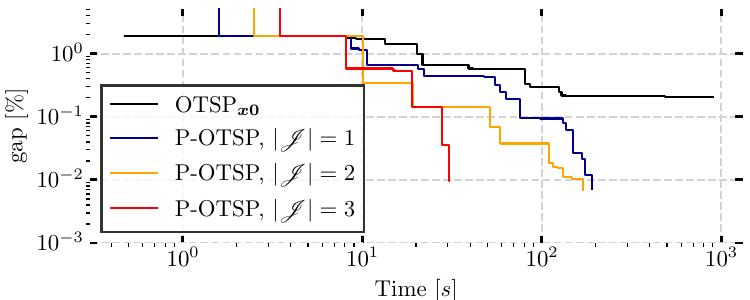}
 \caption{Here, the solution process utilizing varying numbers of parallel ranks are compared, on the 1354\_pegase test case. Every marker denotes obtaining an integer-feasible solution. We use $\otsp_{\xn}$ as a baseline and compare it to P-OTSP with 1, 2, and 3 parallel worker processors. We obtain significant improvements.}
 \label{fig:phtc}
\end{figure}

\begin{figure}[h]
 \centering
 \includegraphics[width=\linewidth]{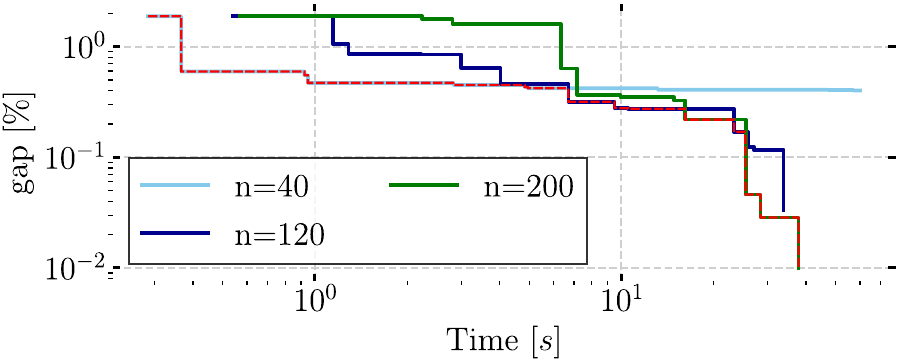}
 \caption{It can be observed, that the solution quality and the time it takes to obtain them is sensitive to $n$, just like in Figure \ref{fig:compHeur}. Whilst small values $n$ result in quick improvements, larger values yield close-to-optimal ones. With $\lvert\J\rvert=3$ we solve all 3 subproblems and the full \otsp simultaneously, sending all solutions immediately to the main optimizer instance on the root process. The red path shows an approximation of the progress our parallel algorithm makes.}
 \label{fig:approx}
\end{figure}

Figure \ref{fig:phtc} impressively shows, that performance, in terms of solution speed and solution quality, improves as more parallel solver instances are utilized. 
The total number of cores (eight) is the same for all four experiments with different parallel architectures. Our parallel algorithm shows to utilize those more efficiently when compared to the stock solver. 
While $\otsp_{\xn}$ cannot find an optimal solution within the time limit, P-OTSP ($\lvert \J \rvert=3$) finds an optimal solution within 30 s. This means more than 6 times lower computations times when compared to P-OTSP ($\lvert \J \rvert=1$ and $\lvert \J \rvert=2$) or more than 30 times lower computation times when compared to $\otsp_{\xn}$.
To outline the source of improvements better, we revisit an adjusted version of Figure \ref{fig:sc1}, as Figure \ref{fig:approx}.

\subsection{Determining AC-feasibility of a topology $\Ea$}

Using the methodology above, many high-quality solutions for \ModelOne can be produced. This can be confirmed visually in Figure \ref{fig:compHeur}, where every marker denotes a feasible solution of the form $(p, f, \theta, x)^{i}$. Here, $x^{i}$ conforms to a set $\Ea_{i}$ by applying \eqref{eq:construction1} in reverse, i.e.
\begin{equation}
\Ea_{i} = \{e\in\E\mid x^{i}_{e}=1\}.\label{eq:ea}
\end{equation}
The previously obtained results are of little significance if no AC-feasible solutions could be found for any of the derived topologies. As \ModelOne is based on the linear DC approximation, we have no guarantees that feasible ACOPF solutions exit.
\newline
An adjusted version of physical AC power flow equations in the polar form is stated as \ModelFive, following \cite{Cain2012a}. 
Additionally to the notation introduced in Table \ref{tab:notation}, $\lvert\bld{v}\rvert$ denote voltage magnitudes and $q$ denote reactive power generation. We, furthermore, consider not only susceptances $b$ but reactances $g$ as well. Lastly, $\bld{f}^{P}_g,~\bld{f}^{Q}_g$ denote cost functions for the real and reactive power output of generators. Considering line switching states like in \ModelTwo, we obtain \ModelFive.

\begin{model}[!h]
\caption{Polar ACOPF: \textsc{ACOPF} ($\Ea$) \hfill [NLP]}
\label{Model5}
\small
\begin{subequations}
\begin{IEEEeqnarray}{r'l}    
    \min z_5 = \sum_{\G}\big[\bld{f}^{P}_{g}(p_{g}) + \bld{f}^{Q}_{g}(q_{g})\big], & \text{{s.t.:}} \\
    \sum_{\E_{v}\cap\Ea} \lvert\bld{v}\rvert_{v^{f}_{e}} \lvert\bld{v}\rvert_{v^{t}_{e}} \big[ g_{e}\text{cos}(\theta_{e}) + b_{e}\text{sin}(\theta_{e}) \big]=~&\nonumber\\ d^{P}_{v} - \sum_{\G_{v}} p_{g},&~\forall~v \in \V\label{eq:acblncp}\\
    \sum_{\E_{v}\cap\Ea} \lvert\bld{v}\rvert_{v^{f}_{e}} \lvert\bld{v}\rvert_{v^{t}_{e}} \big[ g_{e}\text{sin}(\theta_{e}) - b_{e}\text{cos}(\theta_{e}) \big]=~&\nonumber\\ d^{Q}_{v} - \sum_{\G_{v}} q_{g},&~\forall~v \in \V\\
    \underline{q}_{g}\leq q_{g} \leq \overline{q}_{g},&~\forall~g \in \G\\
    \underline{p}_{g}\leq p_{g} \leq \overline{p}_{g},&~\forall~g \in \G\\
    \underline{\theta}_{v}\leq \theta_{v} \leq \overline{\theta}_{v},&~\forall~v \in \V\\
    \lvert\underline{\bld{v}}\rvert_{v}\leq \lvert\bld{v}\rvert \leq \lvert\overline{\bld{v}}\rvert_{v},&~\forall~v \in \V\label{eq:acvmag}
   \end{IEEEeqnarray}
\end{subequations}
\end{model}

First, we introduce a definition of AC-feasibility for a topology $\Ea$ as Definition \ref{def:acfeas}.
\begin{definition}
\label{def:acfeas}
    A topology $\Ea$ is called AC-feasible, if $\exists~ (\lvert\bld{v}\rvert,~\theta,~p,~q)$ satisfying \eqref{eq:acblncp} - \eqref{eq:acvmag}. 
\end{definition}

We confirm AC-feasibility, according to Definition \ref{def:acfeas}, if a solution to \ModelFive can be found.
To solve \ModelFive in the following, MATLAB 2018b \cite{matlab}, MATPOWER 8.0 \cite{Zimmermann2011} and the MATPOWER interior point solver (MIPS)\cite{Wang2007}, version 1.4, are used.
\subsubsection{Repair algorithm for topologies $\Ea_{i}$}\label{sec:repair} Initially, no feasible solutions can be found for \ModelFive on any of the computed topologies, other than for the trivial case, where $x=\bld{1}$. One reason is isolated buses, which are common in solutions for \ModelOne, especially if $d_{v} = 0.0~\text{MW}$, as is the case for buses 87 and 111 in the IEEE 118-bus case. While this could be considered in \ModelOne directly using an additional set of constraints (which makes the solution process more expensive), we use a post-optimization algorithm\footnote{The code is available here: \url{https://github.com/antonhinneck/acfeasibility}.}. 
The procedure is quickly summarized as follows: 
\begin{itemize}
    \item Identify isolated buses \& disconnected groups of buses
    \item Reconnect isolated buses \& disconnected bus groups
    \item Try to solve \ModelFive on the resulting topology
\end{itemize}  

\subsubsection{AC-feasibility results}
\begin{figure}[h]
 \centering
 \includegraphics[width=\linewidth]{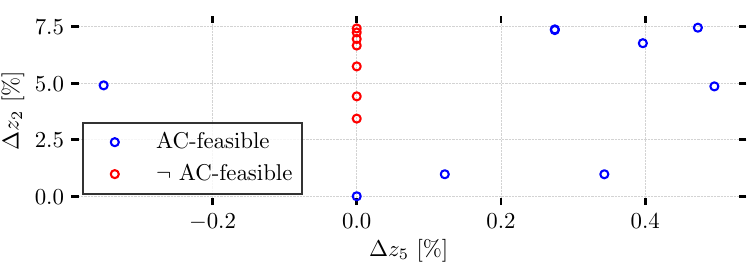}
 \caption{All solutions obtained for the IEEE 118-bus case were tested in \ModelTwo and \ModelFive alike. If a solution was found for a given topology for \ModelFive, we saved its objective value. Infeasible topologies are mapped to 0 \% improvement. It can be observed that about half of the DC-feasible solutions are feasible in \ModelFive. Furthermore, given that a topology is feasible in both problems, improvements in \ModelTwo seem to hold in \ModelFive.}
 \label{fig:acfeas}
\end{figure}

Compared to the original topologies, applying the described repair algorithm results in a much higher success rate. Out of 16 solutions for the 118-bus case, we can confirm AC-feasibility on 9 of them, solving \ModelFive. No convergence, does not mean, however, that the model is infeasible. MIPS could potentially converge given a different starting point for example. The results obtained for the 118-bus case are displayed in Figure \ref{fig:acfeas}. All red circles represent DC-feasible topologies that are not AC-feasible, while blue dots represent topologies that are both AC- and DC-feasible. The blue circle at (0, 0) denotes the worst solution found by Gurobi.

Being able to quickly obtain solutions for \ModelOne using the methods outlined in this paper results in many candidate solutions. Utilizing the proposed repair algorithm, many AC-feasible topologies can be identified quickly. Using our repair algorithm described in Section \ref{sec:repair}, we address common reasons for topologies determined by \ModelOne being infeasible in \ModelFive. Additional adjustments might increase effectiveness and help to derive higher-quality, AC-feasible solutions. This repair algorithm can be seen as a heuristic for transmission switching itself. It, alongside different linear approximations to derive topologies, will be subject to further research.
  \label{sec:results}


\section{Conclusion}
  \label{sec:conclusion}
  In this paper, we proposed a parallel framework to solve combinatorial optimization problems and studied its effectiveness on the optimal transmission switching problem, on 14 case studies. We furthermore, study performance improvements for an increased number of parallel processes, assessing the methods' scalability. The results show that our parallel approach (P-OTSP) can efficiently harness the advantages of both, solving the full \otsp model formulation and heuristics likewise. We lastly provide evidence that the determined topologies can be converted into AC-feasible solutions correcting for common reasons of infeasibilities.
A parallel implementation is advantageous over merely trying to generate improvements using heuristics, because the lower bound on the full problem is available during run time, whereby optimality can be assessed.
The proposed parallel architecture is compatible with any existing MILP solver, such as Gurobi, CPLEX, Xpress, and SCIP - to name a few.
Problem-specific switching criteria like \lpsc have shown very effective when it is used in heuristics that run in parallel to the main \otsp solution procedure. Problem-specific switching criterion like \lpsc has shown effective when used in a heuristic method that runs parallel to the main \otsp solution procedure.  
The proposed parallel method shows better performance than B\&C algorithms of off-the-shelve solvers.

The optimality \textit{gap} reported for the 14 case studies, shows that, except for the 6470\_rte case, solutions within 0.86\% of the full \otsp's globally optimal objective value's can be obtained in less than 15 minutes. 
Even if an optimal solution cannot be obtained within the time limit, significantly better solutions are available.

Lastly, a high degree of parallelization can reduce computation times by a factor of 30. 
The framework proposed is not limited to any particular heuristic or problem. Instead of MIP solvers, heuristics could be run in parallel.
A direction for future work is the application of the proposed methods to different discrete problems in power systems.
Moreover, generating a priority list $\lp$ could be done using different rules or machine learning algorithms.
The potential is huge and goes beyond the optimal transmission switching problem. Many combinatorial problems are repeatedly solved in practice, domain-specific knowledge, data-driven algorithms, or machine learning could help to accelerate exact algorithms.
A final implication is this; approaching transmission switching in the presented manner might provide strategic implications for practitioners.
The solutions obtained not only provide improvements but also theoretical guarantees as to how much improvement could still be achieved. 
It poses a stronger basis for future decisions than cost reduction alone.

\section*{Acknowledgment}

This work was funded by the Skolkovo Institute of Science and Technology as a part of the Skoltech NGP Program (Skoltech-MIT joint project). 
This paper was prepared as a part of the AMPaC Megagrant project supported by Skoltech and The Ministry of Education and Science of Russian Federation, Grant Agreement No 075-10-2021-067, Grant identification code 000000S707521QJX0002.

\ifCLASSOPTIONcaptionsoff
  \newpage
\fi



\bibliographystyle{IEEEtran}
\bibliography{bib}{}
%


%

\begin{IEEEbiography}[{\includegraphics[width=1in,height=1.25in,clip,keepaspectratio]{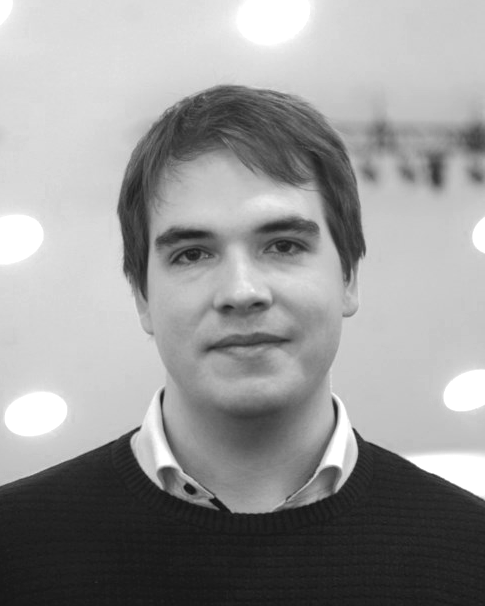}}]{Anton Hinneck}
received his B.S. and M.S. degrees at Berlin Institute of Technology in Industrial Engineering and Management in 2015 and 2018 respectively. He is currently working towards his Ph.D. degree at the Skolkovo Institute of Science and Technology at the Center for Energy Science and Technology. His research interests include power systems, electricity markets, uncertainty and mixed-integer optimization.
\end{IEEEbiography}

\begin{IEEEbiography}[{\includegraphics[width=1in,height=1.25in,clip,keepaspectratio]{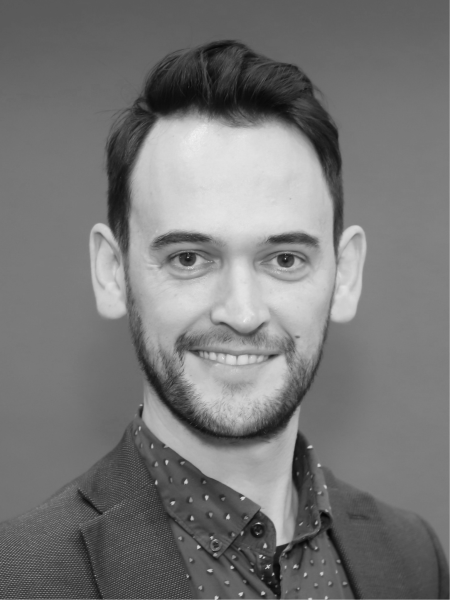}}]{David Pozo}
(S’06–M’13–SM’18) received the B.S. and Ph.D. degrees in electrical engineering from the University of Castilla-La Mancha, Ciudad Real, Spain, in 2006 and 2013, respectively. Since 2017, he has been an Assistant Professor with the Skolkovo Institute of Science and Technology (Skoltech), Moscow, Russia. Prior to Skoltech, he was a Postdoctoral Fellow with the Pontifical Catholic University of Chile and the Pontifical Catholic University of Rio de Janeiro. His research interests include power systems, operations research, uncertainty, Game Theory, electricity markets, and problems of optimization and flexibility of modern power systems. Since 2018, he has been leading the research group on Power Markets Analytics, Computer Science and Optimization, at Skoltech.
\end{IEEEbiography}

\end{document}